\documentclass[
reprint,
 amsmath,amssymb,
 aps,
 notitlepage,
nofootinbib,
superscriptaddress,
float
]{revtex4-1}

\usepackage[detect-all]{siunitx}
\usepackage[export]{adjustbox}
\usepackage{float}
\usepackage{graphicx}
\usepackage{dcolumn}
\usepackage{bm}
\usepackage{eqnarray}

\usepackage[hidelinks]{hyperref} 

\usepackage{placeins}

\usepackage[caption=false]{subfig}

\usepackage{float}

\usepackage{xr}
\begin{document}

\title{Ultrafast X-ray imaging of pulsed plasmas in water}

\author{Christopher Campbell}
\affiliation{Department of Mechanical Engineering, Texas A\&M University, College Station, TX 77843, USA}
\author{Xin Tang}
\affiliation{Department of Mechanical Engineering, Texas A\&M University, College Station, TX 77843, USA}
\author{Yancey Sechrest}
\affiliation{Los Alamos National Laboratory, Los Alamos, NM 87545, USA}
\author{Kamel Fezzaa}
\affiliation{X-ray Science Division, Advanced Photon Source, Argonne National Laboratory, Argonne, IL 60439, USA.}
\author{Zhehui Wang}\email[Correspondence to: ]{zwang@lanl.gov}
\affiliation{Los Alamos National Laboratory, Los Alamos, NM 87545, USA}
\author{David Staack}\email[Correspondence to: ]{dstaack@tamu.edu}
\affiliation{Department of Mechanical Engineering, Texas A\&M University, College Station, TX 77843, USA}

\date{\today}

\begin{abstract}
Pulsed plasmas in liquids exhibit complex interaction between three phases of matter (liquids, gas, plasmas) and are currently used in a wide range of applications across several fields, however significant knowledge gaps in our understanding of plasma initiation in liquids hinder additional application and control; this area of research currently lacks a comprehensive predictive model.
To aid progress in this area experimentally, here we present the first-known ultrafast (50 ps) X-ray images of pulsed plasma initiation processes in water (+25 kV, 10 ns, 5 mJ), courtesy of the X-ray imaging techniques available at Argonne National Laboratory's Advanced Photon Source (APS), with supporting nanosecond optical imaging and a computational X-ray diffraction model.
These results clearly resolve narrow ($\sim$$10\ \si{\micro\meter}$) low-density plasma channels during initiation timescales typically obscured by optical emission ($<$$100$ ns), a well-known and difficult problem to plasma experiments without access to state-of-the-art X-ray sources such as the APS synchrotron.
Images presented in this work speak to several of the prevailing plasma initiation hypotheses, supporting electrostriction and bubble deformation as dominant initiation phenomena.
We also demonstrate the plasma setup used in this work as a cheap ($<$US\$100k), compact, and repeatable benchmark imaging target (29.1 km/s, $1\ \si{\tera\watt/\centi\meter^2}$)  useful for the development of next-generation ultrafast imaging of high-energy-density physics (HEDP), as well as easier integration of HEDP research into synchrotron-enabled facilities.  
\end{abstract}

\maketitle

\begin{figure}
\includegraphics[width=0.9\linewidth,valign=t]{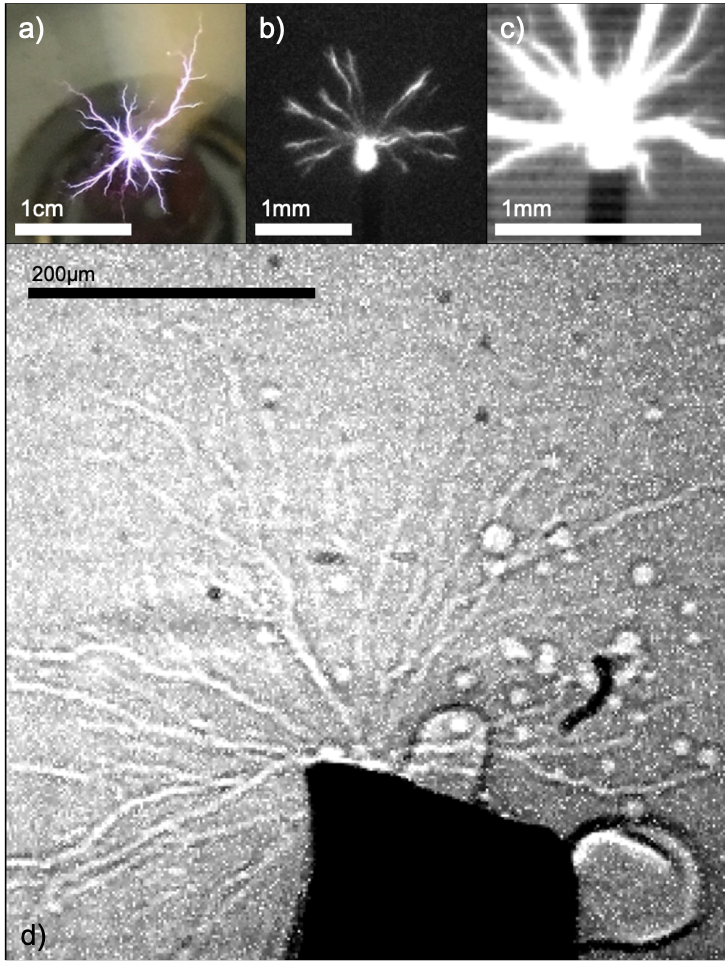}
\caption{The pulsed water plasma of interest to this work imaged with four different methods, in order of decreasing exposure time: 67ms (a), 2.38$\si{\micro\second}$ (b), 10ns (c) and 50ps X-ray (d).}
  \label{fig1}
\end{figure}

\noindent\emph{Introduction --}
In the past two decades, the study of plasma processes in liquids has grown to be a vigorous, broadly interdisciplinary field that has found numerous applications in chemical processing \cite{Hao_2015}, nanomaterial synthesis \cite{Bhattacharyya_2009}, medicine \cite{Hutson_2007}, and many other fields.
Despite the array of applications of plasma processes in liquids, significant knowledge gaps in our understanding of plasmas physics of liquids, and of fundamental physical interactions between the second (liquid) and fourth (plasma) states of matter, persist \cite{Vanraes_2018}.
Historically, research into pulsed plasma initiation phenomena has focused on gas-phase processes, which as a result are relatively well understood \cite{Raizer_GDP,Piel_PlasmaPhysics}.
In contrast, electrical discharges in liquids exhibit a complex multi-phase environment through which the plasma propagates \cite{Chu_2011}, and, consequently, plasma initiation in liquids is still not understood.
Recent effort has focused on investigation into liquid-phase breakdown initiation phenomena, of which there are several prevailing hypotheses regarding the dominant breakdown mechanism  and currently no conclusive theory \cite{Vanraes_2018}.
It has not yet been proven whether electron avalanching is possible in liquid media, therefore several hypotheses require the presence and deformation of preexisting bubbles (possibly sub-microscale) or dissolved gases for streamer propagation to occur \cite{Starikovskiy_2011}.
Others propose the generation of nano-pores via electrostriction as a source of low-density regions required for plasma initiation \cite{Dobrynin_2013}.
Still others suggest that field emission from the electrode tip can locally heat the surrounding water, leading to rapid expansion and a low-density region through which electrons can avalanche \cite{Jones_1995}.

This work employs optical imaging and ultrafast X-ray phase-contrast imaging to provide unprecedented characterization of the plasma initiation mechanism of nanosecond pulsed plasma discharges in water. 
These measurements capture the rapid (29 km/s) expansion of a network of 10-$\si{\micro\meter}$ diameter, low density channels that manifest on timescales on the order of 10 ns.
Taken together, these measurements narrow the field of possible dominant physical processes for nanosecond plasma initiation in liquids to those which rely on nanosecond-timescale physics, supporting electrostriction and bubble deformation while weakening the local field-emission heating hypothesis.

In addition to its utility as a plasma diagnostic, the X-ray imaging in this work represents the introduction of pulsed plasmas in liquids to the field of synchrotron science.
The tabletop setup used in this work is quite portable and relatively cheap ($<$US\$100k), while still providing hypersonic phenomena (29.1 km/s) and high energy densities ($1\ \si{\tera\watt/\centi\meter^2}$) using as little as 5 mJ of plasma energy.
An equivalent plasma device could therefore be easily incorporated as a self-healing benchmark imaging target for diagnostic development at ICF facilities (e.g. National Ignition Facility \cite{Cerjan_2018}), as well at any other facility interested in high-energy-density physics (HEDP) and fast phenomena \cite{Lee_2012, Wildeman_2017, Wang_2008}.
This partnership between plasma physics and synchrotron science opens new avenues for interrogating sub-nanosecond plasma dynamics in liquids, as well as interrogating and validating HEDP science at synchrotron-enabled facilities.

\begin{figure}[h]
\includegraphics[width=0.9\linewidth,valign=c]{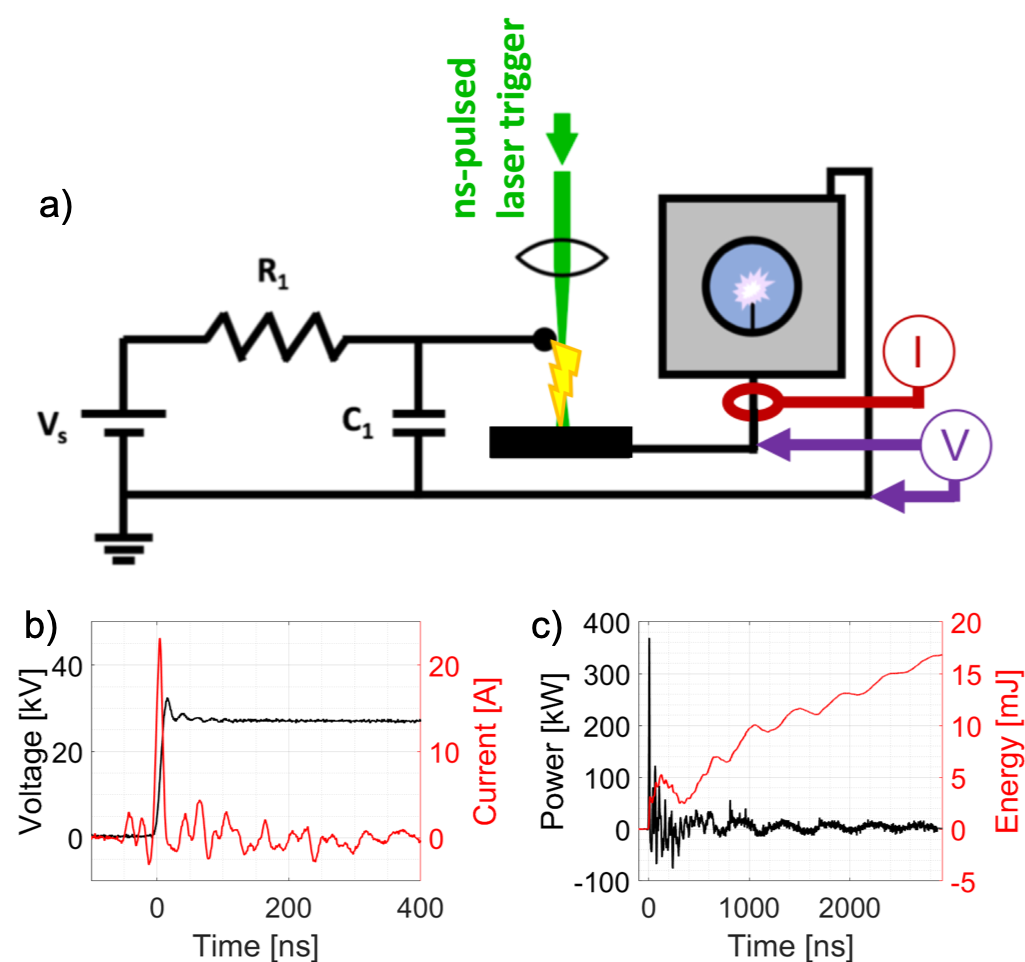}
\caption{\label{fig:electrical}
Diagram of the laser-triggered driving circuit (a), with voltage and current traces for a typical event (b).  
Power and energy calculated from direct integration (c).
$R_1 = 20\ \si{\mega\ohm}$, $C_1 = 1\ \si{\nano\farad}$, and $V_b = +25\ \si{\kilo\volt}$.
During this event a total of 242 mJ was dissipated: 225 mJ ($93\%$) was dissipated by the air spark switch, 5 mJ ($2\%$) contributed to plasma generation, and the remaining 12 mJ ($5\%$) was lost via long-timescale heating and electrolysis.
The resulting plasma current peak had a FWHM of 12 ns and a peak of 20 A.
}
\end{figure}

\medbreak\noindent\emph{Phenomena of Interest --}
This work focuses on the pulsed plasma shown in Figure \ref{fig1}, generated using the driving circuit shown in Figure \ref{fig:electrical}a.
When applying a positive high-voltage pulse to a 200-$\si{\micro\meter}$-diameter tungsten electrode submerged in distilled water (filtered to a maximum particle size of $0.2\ \si{\micro\meter}$, with conductivity of roughly $0.5\ \si{\micro S/\centi\meter}$), a branched-structure plasma forms near sharp-contour regions of the electrode tip, radiating outward at hypersonic velocities.
By triggering an air spark gap with a nanosecond laser pulse (similar to \cite{Lopes_2003}), repeatable time-resolved diagnostics are possible.
While a nanosecond-pulsed plasma can occur in a more conductive liquid \cite{Xiao_2014}, the branched structures of interest to this work require low conductivity, like that of distilled water or lower.
See Section SM.\ref{setupspecs} for a more detailed description of the experimental setups and methods used in this work.

\begin{figure} 
\includegraphics[width=\linewidth]{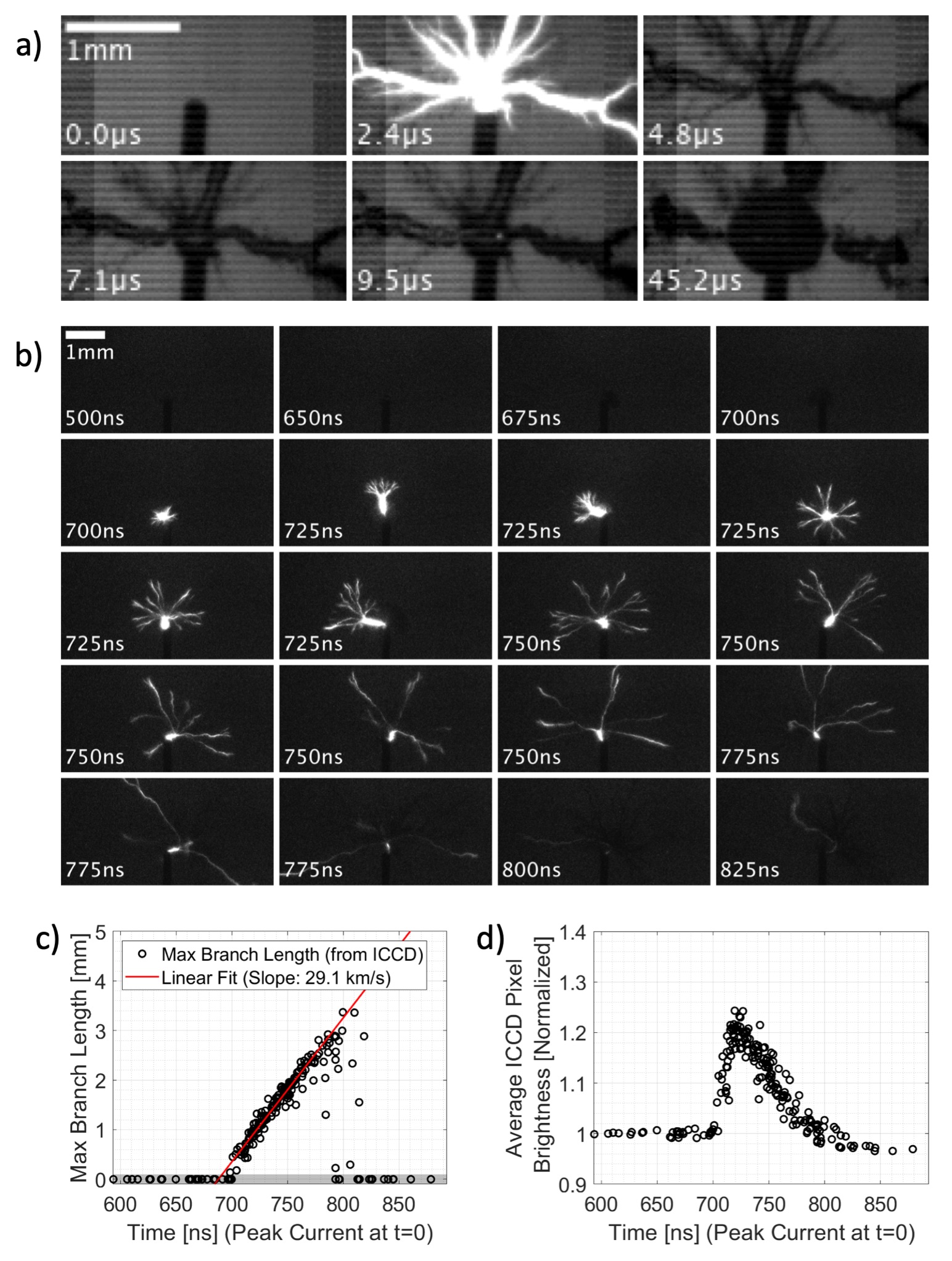}
\caption{A series of video frames from high-speed (2.38 $\si{\micro\second}$/frame) backlit photography of a single water discharge event (a, Supplementary Video 1).
Collection of backlit fast-exposure (10ns) ICCD images from many disparate events, sorted by camera delay relative to time of event as measured by peak PMT signal, with $t=0$ defined to be the average time of peak current (b, Supplementary Video 2). 
Images from (b) were used to generate plots of plasma channel length (c) and average frame brightness (d) vs. time.}
\label{fig:hispeed}
\end{figure}

\medbreak\noindent\emph{Optical Imaging Results --}
Using a high-speed video camera, the plasma of interest was imaged with backlighting at 420 kfps (2.38 $\si{\micro\second}$/frame) and a resolution of 20 $\si{\micro\meter}$/pixel, shown in Figure \ref{fig:hispeed}a and Supplementary Video 1.
Light-emitting plasma channels propagate across the full field of view within the first captured frame of the event, implying a lower bound of $~630$ m/s for propagation speed.
Optical emission and plasma energy deposition has ceased by the next frame, and the resulting bubble begins to evolve and become more spherical over the next 50 $\si{\micro\second}$.

To better interrogate plasma-timescale processes without requiring ultra-high-frame-rate imaging of a single event, we can take advantage of the fact that this plasma process and experimental setup  is well-timed.
By triggering the event relative to the shutter of a nanosecond-gated camera and logging image timing signals, single event images were sorted according to exposure delay relative to each event.
This results in a constructed ``video'' of water plasma behavior at timescales not achievable via single event high-frame-rate imaging, shown in Figure \ref{fig:hispeed}b and Supplementary Video 2.
The plasma (identified by the light-emitting region) initiates near the electrode tip and propagates outward with a hemispherical bush/branching structure over approximately 100 ns, with the longest branches extending almost $4\ \si{\micro\meter}$ away from the electrode before extinguishing.
Outward propagation speed is therefore inferred to be $29.1\ \si{km/s}$ (Mach 19.7 in water),  estimated from the linear trend shown in Figure \ref{fig:hispeed}c.
This propagation speed agrees with prior literature \cite{Ceccato_2010} where the initiating plasma extends out in front of an expanding shockwave.
Such a high Mach number suggests that the dominant initiation process is not limited by ambient sound speed, undermining plasma initiation hypotheses which rely on slower processes such as Joule heating or electrolysis.
While slower processes may still be significant during longer timescales, they do not appear to drive the channel tip.
However, the significant optical emission obscures multi-phase phenomena needed for more informed discussion of initiation processes.
This issue, along with diffraction-limited resolution at micron scales, prompted our interest in the fast X-ray techniques available at the Advanced Photon Source (APS).

\begin{figure*}
\centering
  \includegraphics[width=\linewidth]{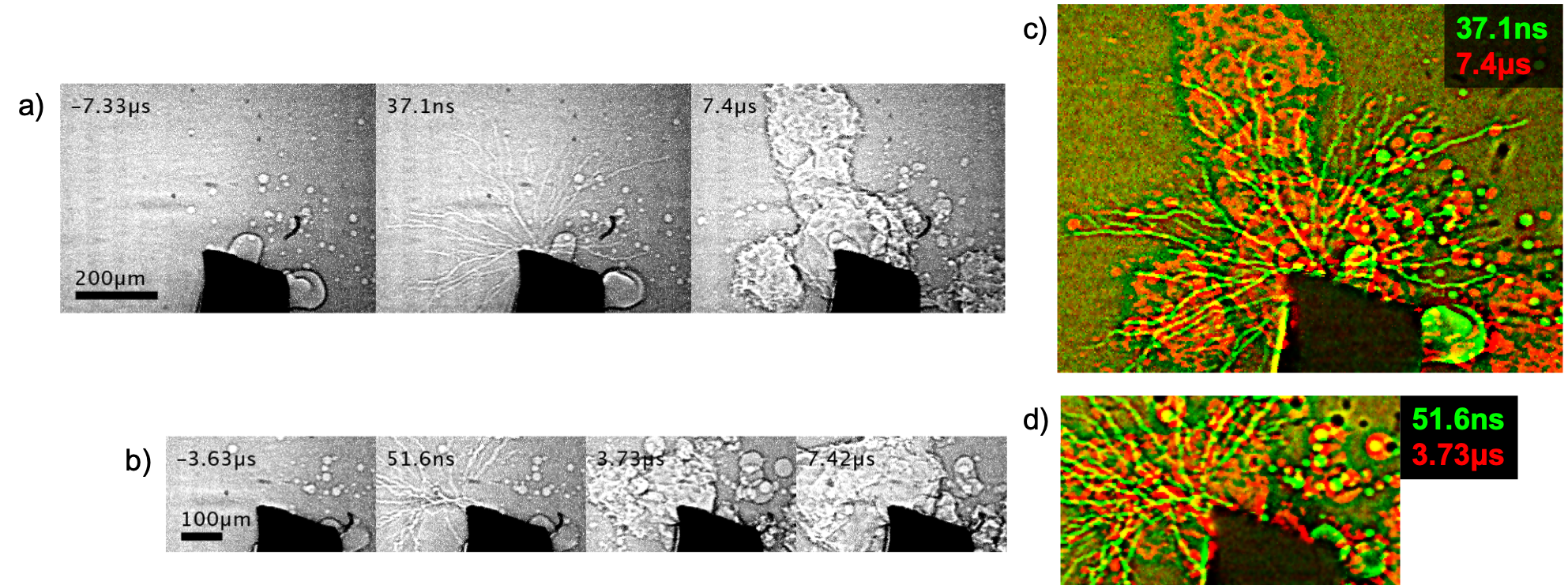}
\caption{
 Frames from ultrafast X-ray imaging videos of two pulsed water plasma events (a and b), with timestamps annotated relative to peak plasma current as measured through the positive electrode.
 Frame rates are 136 kfps (a) and 272 kfps (b), or 7.37 and 3.69 $\si{\micro\second}$/frame respectively.
 Resolution is 2~$\si{\micro\meter}$/pixel.
 Note the evolution from pre-discharge to small-diameter long plasma channels to large-diameter cavitation and expansion, as emphasized in (c) and (d) with two-frame composite images from Figures (a) and (b), respectively.
See Section SM.\ref{moreevents} for additional multi-frame videos of single events, all of which show correlation in position between the small-diameter channels and larger-diameter bubbles.
 Figure \ref{fig:xraysingleevent1}a corresponds to Supplementary Video 3.
 }
  \label{fig:xraysingleevent1}
\end{figure*}

\medbreak\noindent\emph{X-ray Imaging at APS 32-ID-B --}
To integrate our setup into the existing X-ray imaging facilities at the APS 32-ID-B beamline lab (see Figures SM\ref{sm-setup} and SM\ref{fig:imagingsetup}), several modifications to the water discharge cell and the high-voltage circuit were necessary.
Since the X-ray attenuation length of water is 2.2 cm for the X-ray energies relevant to this work (24.3 keV), a water discharge cell was designed which minimizes the intra-device path length of the X-ray beam to 2 mm.

See Figure \ref{fig:xraysingleevent1} and Supplementary Video 3 for typical X-ray imaging results from separate events, imaged using 24.3-keV X-rays from the APS synchrotron source during a hybrid fill operating mode \cite{APSparams}.
The X-ray setup at APS 32-ID-B was configured in a phase-contrast imaging mode, which exploits relative phase delay of the X-ray beam induced by objects on the sample plane with large density gradients \cite{Fezzaa_2008,Diemoz_2013,Endrizzi_2017}.
The gas-liquid boundary of a plasma-induced bubble is therefore easily resolvable in these phase-contrast images with negligible motion blur.
See Figures SM\ref{fig:xray-mf1}--SM\ref{fig:xray-mf5} for more X-ray imaging examples of single-events.

Before any quantitative analysis of these X-ray images, there are important qualitative observations to be made.
In the first post-initiation frame of each imaged event (such as the second frames of Figures \ref{fig:xraysingleevent1}a and \ref{fig:xraysingleevent1}b), we observe the outward propagation of channel-like features at timescales much faster than the inter-frame period ($3.69\ \si{\micro\second}$), with presumed dynamic behavior within the first 100 ns after initiation.
The X-ray videos presented here exclusively show plasma channels which traverse the full field of view within a single inter-frame duration, and the stochasticity of this plasma process hinders insight from multi-event comparison.
That being said, this work is still an important result in itself; plasma channel evolution during periods of significant plasma optical emission has not been previously observed.

This phase-contrast imaging method amplifies the effect of discontinuities in density such as surfaces, therefore we interpret these features to be the liquid-gas interfaces of a low-density channel propagating with or near the optical emission front.
We also observe correlation in location and propagation direction between these narrow ($\sim 10\ \si{\micro\meter}$) channels and the larger-diameter non-spherical bubbles visible in the subsequent frame $3.69\ \si{\micro\second}$ later, emphasized by the two-frame composite image in Figure \ref{fig:xraysingleevent1}c.
Depending on the unequal distribution of energy deposition, the thin plasma branches will either expand into a large diameter bubble (such as the branches at 8 o'clock and 11 o'clock positions in Figure \ref{fig:xraysingleevent1}), or will collapse back down to a train of small spherical bubbles (such as the branch at the 9 o'clock position in Figure \ref{fig:xraysingleevent1}).

Occurring next after bubble expansion, the long-lived abnormal bubble shape visible in each case persists for almost $50\ \si{\micro\second}$ after initiation (see Figures SM\ref{fig:xray-mf1}-SM\ref{fig:xray-mf5}). 
Based on this, we conclude that significant charge resides near the bubble surface for tens of microseconds after initiation, resulting in long-lived local Coulomb forces after plasma energy deposition.
This is consistent with the characteristic charge relaxation time expected for ambient water, $\tau = \sigma/\varepsilon \approx 3.6\ \si{\micro\second}$.
The observed sharp-contour bubbles suggest low local surface tension, caused by the Lippmann effect acting at the locally-charged interface.
Though the resulting bubble shape is quite striking, processes which dominate at these longer ($\si{\micro\second}$) timescales are not necessarily responsible for initiation.

It is also important to note the plasma-independent continuous formation of bubbles visible within the region of interest.
As previously discussed the water was sufficiently filtered ($<0.2\ \si{\micro\meter})$, ruling out the possibility of particle contamination.
Future work may investigate the effect of degassing on the production of these bubbles, however in this case we attribute this bubble production to water radiolysis and electrode heating, both induced by high X-ray power ($21\ \si{\watt/\milli\meter^2}$ \cite{XOP}).
Future work may investigate the effect of degassing on the production of these bubbles, however radiolysis and heating due to high X-ray power is the most likely cause in these results.
X-ray-induced bubble production affects the timing of nanosecond plasma initiation processes in water, however the rapidity of such processes appears to be unaffected during the timescales imaged here.

\medbreak\noindent\emph{Computational Model --}
Unlike a near-field X-ray image which can be easily converted to line density, a phase contrast image requires a more rigorous approach to quantitatively analyze; in general a two-dimensional Fresnel-Kirchhoff diffraction integral is required \cite{Pedrotti3}. 
For this particular case we can take advantage of both the small scattering angle of the X-ray beam after the sample plane and the cylindrical geometry of a plasma channel to compose a simplified integral:

\begin{flalign}
\label{modeleq}g_\text{out}(x') &= \frac{e^{2 \pi i z/\lambda}}{\sqrt{i \lambda z}} \int_{-\infty}^{\infty} g_\text{in}(x) e^{\frac{i \pi}{\lambda z} (x' - x)^2} dx \\
I_\text{detector}(x') &= g_\text{out}(x')g^*_\text{out}(x')
\end{flalign}
\vspace{1pt}

\noindent where $x$ is lateral position relative to the plasma channel at the sample plane and $x'$ is lateral position at the detector plane. 
The complex-valued functions $g_\text{in}(x)$ and $g_\text{out}(x')$ describe the electric field at the sample plane and detector plane, respectively.
We assume the geometry shown in Figure SM\ref{fig:imagingsetup}, which consists of a simple cylinder of specific gravity $SG_\text{vap}$ surrounded by ambient water ($SG_\text{liq} = 1$).
The term $g_\text{in}(x)$ is therefore defined as follows for a plasma channel of radius $R = \frac{1}{2}d_\text{channel}$ in a water cell of width $w_\text{liq}$:

\begin{flalign}
g_\text{in}(x) = 
\begin{cases}
e^{\big[\frac{-2 i \pi}{\lambda}\tilde{n}_\text{water,liq}\cdot w_\text{liq}\big]}, |x|>R \\
e^{\big[\frac{-2 i \pi}{\lambda}\tilde{n}_\text{water,liq}\cdot(w_\text{liq}-2\sqrt{R^2-x^2})\big]} \\ \quad \cdot e^{\big[\frac{-2 i \pi}{\lambda}\tilde{n}_\text{water,vap}\cdot(2\sqrt{R^2-x^2})\big]}, |x| \leq R 
\end{cases}
\end{flalign}

\noindent where the complex refractive index $\tilde{n} = 1-\delta-i\beta$ is linearly dependent on water density: $\delta_\text{vap} = SG_\text{vap}\delta_\text{liq}$ and $\beta_\text{vap} = SG_\text{vap}\beta_\text{liq}$, with $\delta_\text{liq}$ and $\beta_\text{liq}$ taken from literature \cite{Henke_1993}.
Derivation of this model is described in further detail in Section SM.\ref{sm-modelderivation}, which builds off of the general form from \cite{MITocw_fresnel} and is equivalent to work by Snigirev on phase-contrast imaging of cylindrical samples \cite{Snigirev_1995}.

\begin{figure}
\includegraphics[width=\linewidth]{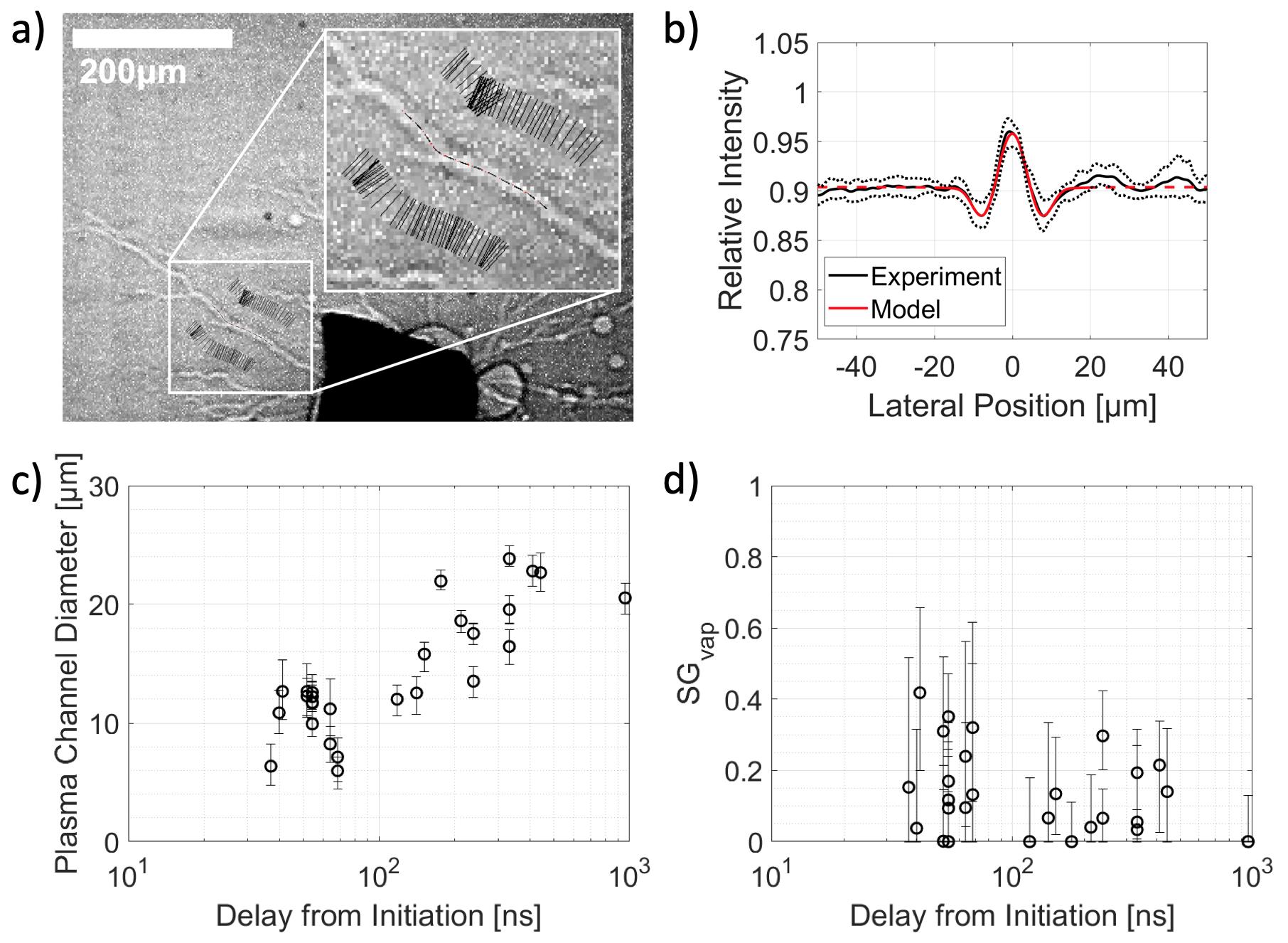}
\caption{
  Illustration of a typical plasma channel cutline interpolation from an X-ray image (a), typical diffraction model result after optimization compared with the average experimental cutline (b). 
  Intensity measured relative to incident X-ray intensity.
  Dotted lines represent first and third quartiles of the cutline distribution for each lateral position.
  For this particular result, the channel diameter and specific gravity were found to be $11.7^{+1.5}_{-1.3}\ \si{\micro\meter}$ and $0.09^{+0.17}_{-0.09}$, respectively.
Computational results for 31 interpolated experimental cutlines are summarized in (c) $d_\text{channel}$ vs. $t$  and (d) $SG_\text{vap}$ vs. $t$ plots, with time measured relative to water plasma initiation.
Note the general trend that longer times after initiation correlate with larger-diameter plasma channels and lower specific gravities, on timescales much faster than the X-ray frame rate.
See Figure SM\ref{fig:sensitivity} for discussion on uncertainty.
}
\label{fig:model1}
\end{figure}

By numerically integrating Equation \ref{modeleq}, we can generate a simulated phase-contrast X-ray image.
Parameters needed for this model include the specific gravity of the region within the plasma channel $SG_\text{vap}$, diameter of the channel $d_\text{channel}$, as well as three minor parameters which scale and shift the model.
To convert the X-ray image into a model-comparable form, we interpolate the image onto a series of cutlines oriented perpendicular to a spline estimation of the plasma channel centerline, forming a cutline intensity distribution for each lateral position.
The model is then fit to experimental results to extract estimates of $SG_\text{vap}$ and $d_\text{channel}$ (see Section SM.\ref{sm-modelderivation} for details), as shown in Figure \ref{fig:model1}.

From this fitted model, $SG_\text{vap}$ and $d_\text{channel}$ of the plasma channel can be estimated for a particular interpolated cutline, as shown in Figure  \ref{fig:model1}.
The first evident conclusion from these results is that these plasma channels are exclusively low-density phenomena which propagate at speeds comparable to those calculated from the optical emission region (see Figure \ref{fig:hispeed}).
This supports initiation hypotheses which require the initial generation of lower-density voids for plasma propagation, such as electrostriction and deformation of preexisting bubbles.
It is also important to consider the effect of photon-electron interaction in this environment due to high electron density (on the order of $10^{18}\ \si{\per\cubic\centi\meter}$), however in this case we do not believe that this phenomena has significantly affected these X-ray imaging results, due to the small scattering angle used for imaging.
 A total of 31 model fits were produced for a variety of selected cutlines at different imaging delays relative to peak current and different distances from the electrode tip (as shown in Figures \ref{fig:model1}c, \ref{fig:model1}d, and SM\ref{modelsummaryscatter}), collectively revealing a few major trends in the X-ray data; in particular, larger channel diameters (Figure \ref{fig:model1}c) and lower specific gravities (Figure \ref{fig:model1}d) occur with increasing time after initiation.
These results imply that low-density regions visible in X-ray occur at comparable or possibly earlier times (relative to initiation) than those of the light-emitting region, however further investigation is necessary to show this conclusively.

\medbreak\noindent\emph{Conclusion --}
The above work presents provides insight into the particular mechanism of breakdown in nanosecond-pulsed plasmas in liquids. 
Using a combination of X-ray and optical methods, we have resolved narrow low-density plasma channels within the streamer head which evolve at speeds comparable to those of the light-emitting region.
To the best of our knowledge, phase-contrast X-ray imaging has not been previously explored as a diagnostic for such plasmas.
The resulting superior imaging resolution for this well-timed process and insensitivity to plasma optical emission provides insight about dominant plasma initiation mechanism hypotheses (supporting electrostriction and bubble deformation while weakening local field-emission heating), and encourages further interaction between the fields of plasma physics and synchrotron science, both as a phenomenon of primary research interest as well as a tabletop self-healing HEDP imaging target.
Future work will include additional plasma imaging experiments at APS with the added experience gained from results reported here, such as the use of sharper electrode tips ($<10\ \si{\micro\meter}$) for better initiation consistency and higher X-ray imaging frame rates to reveal dynamic processes for single events at timescales less than 150 ns (frame rates of 6.67 MHz are expected during future experimental campaigns, compared to 272 kHz in this work).
Continued progress will further contribute to better understanding of plasma initiation in liquids, and overall increased interdisciplinary work between the fields of plasma physics, HEDP, and synchrotron science.

\medbreak\noindent\emph{Acknowledgements --}
Special thanks to the High-Speed Imaging Team at Los Alamos National Laboratory for their collaborative efforts and financial support during these experiments, and to the staff at APS 32-ID-B for their time and expertise. 
Los Alamos National Laboratory (LANL) work is supported  through  Triad  National  Security,  LLC  (`Triad') by U.S. Department of Energy  (DOE)/ NNSA, by MaRIE Technology Maturation fund and C2 program.
This research used resources of the Advanced Photon Source, a U.S. Department of Energy (DOE) Office of Science User Facility operated for the DOE Office of Science by Argonne National Laboratory under Contract No. DE-AC02-06CH11357.

\bibliography{mybib}
\bibliographystyle{ieeetr}

\end{document}